\documentclass[aps,pra,twocolumn,superscriptaddress,showpacs]{revtex4}

\usepackage{graphicx}
\usepackage{color}
\usepackage[hypertex,unicode=true]{hyperref}

\hypersetup{
        colorlinks=true,
        citecolor=blue,
        linkcolor=black,
        urlcolor=blue}

\newcommand{\half}{\ensuremath{\frac{1}{2}}}
\newcommand{\halfl}{\ensuremath{{\scriptstyle \frac{1}{2}}}}
\newcommand{\real}{\text{Re}}
\newcommand{\imag}{\text{Im}}

\newcommand{\un}   [1]{\ensuremath{\,\mathrm{#1}}}
\newcommand{\unit} [1]{\un{#1}}

\newcommand{\intd}    {\ensuremath{\,\mathrm{d }}}

\newcommand{\text} [1]{\ensuremath { \mathrm {#1} }}

\newcommand{\pder} [2]{\ensuremath{\frac{\partial #1}{\partial #2}}}
\newcommand{\pderl}[2]{\ensuremath{\partial #1/\partial #2}}

\newcommand{\tderl}[2]{\ensuremath{\text{d} #1/\text{d} #2}}

\newcommand{\gom}  {\ensuremath{g_{\mathrm{OM}}}}
\newcommand{\com}  {\ensuremath{c_{\mathrm{OM}}}}
\newcommand{\nmax} {\ensuremath {n_{\mathrm{max}}}}
\newcommand{\uzpm}   {\ensuremath {u_{\mathrm{zpm}}}}

\begin{document}


\title{Backaction limits on self-sustained optomechanical oscillations}

\author{M.~Poot}
\email{menno.poot@yale.edu} \affiliation{Department of
Electrical Engineering, Yale University, New Haven, CT 06520,
USA}

\author{K.~Y.~Fong}
\affiliation{Department of Electrical Engineering, Yale
University, New Haven, CT 06520, USA}

\author{M.~Bagheri}
\affiliation{Department of Electrical Engineering, Yale
University, New Haven, CT 06520, USA}

\author{W.~H.~P.~Pernice}
\affiliation{Department of Electrical Engineering, Yale
University, New Haven, CT 06520, USA}

\author{H.~X.~Tang}
\email{hong.tang@yale.edu} \affiliation{Department of
Electrical Engineering, Yale University, New Haven, CT 06520,
USA}

\date{\today}

\begin{abstract}
The maximum amplitude of mechanical oscillators coupled to
optical cavities are studied both analytically and numerically.
The optical backaction on the resonator enables self-sustained
oscillations whose limit cycle is set by the dynamic range of
the cavity. The maximum attainable amplitude and the phonon
generation quantum efficiency of the backaction process are
studied for both unresolved and resolved cavities. Quantum
efficiencies far exceeding one are found in the resolved
sideband regime where the amplitude is low. On the other hand
the maximum amplitude is found in the unresolved system.
Finally, the role of mechanical nonlinearities is addressed.
\end{abstract}

\pacs{
85.85.+j, 
42.50.Wk  
}

\maketitle

\section{Introduction}
Optomechanics is a rapidly growing field in which the
interaction between optical cavities and mechanical resonators
is studied, see Refs.
\cite{kippenberg_science_optomechanics_overview,
marquardt_physics_optomechanics_overview,
favero_natphot_overview, poot_physrep_quantum_regime} for
recent reviews. A lot of interest lies in fundamental questions
such as the role of gravity in quantum decoherence, the
ultimate limits on position detection, and backaction evasion.
These questions might be answered by cooling the mechanical
resonator to its ground state. Only very recently this has been
achieved using the photon pressure in microwave
\cite{teufel_nature_groundstate} and optical cavities
\cite{chan_nature_groundstate}. Using red detuned light the
thermal vibrations of the resonator could be cooled below the
zero-point motion, which typically lies in the femtometer range
\cite{poot_physrep_quantum_regime}. On the other hand, for
practical applications, one would like to have amplitudes as
large as possible. This can be achieved with blue detuned
light, which amplifies the motion of the resonator. For
example, we have recently demonstrated a non-volatile
mechanical memory \cite{bagheri_natnano_high_amplitude} and
synchronization of remote mechanical oscillators
\cite{bagheri_synchronization} using the large motion
amplitudes generated by the optical backaction of an on-chip
racetrack cavity. Large amplitude self-sustained oscillations
(SSOs) also enabled the observation of chaotic dynamics
\cite{carmon_PRL_chaos} and the zero-frequency anomaly
\cite{bagheri_natnano_high_amplitude}. Reaching high amplitude
motion is thus important for both technological advances as for
fundamental research. Here, we address the question what
ultimately limits the maximum amplitude of regenerative
oscillations in a cavity-optomechanical system.

\begin{figure}[tb]
\includegraphics{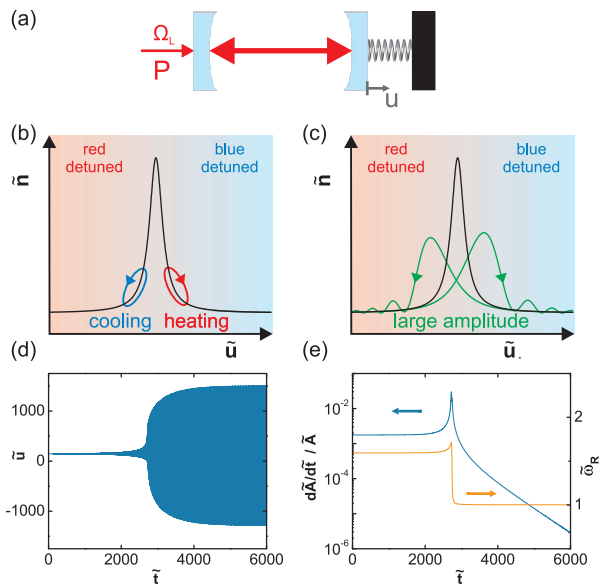}%
\caption{(Color online) (a) Schematic representation of a generic optomechanical system.
Laser light with power $P$ enters the Fabry-P\'erot cavity through the left mirror
and bounces back and forth between the two mirrors. The right mirror
acts as a mechanical resonator.
(b) Optical backaction in the small amplitude limit. A delay
between the displacement $\tilde u$ and the change in the photon occupation
$\tilde n$ due to the finite cavity lifetime $\tilde \kappa^{-1}$ leads to ellipsoidal
trajectories whose area is proportional to the work done on the resonator.
Negative (positive) work leads to cooling (heating) of the resonator in the red (blue) detuned
region.
(c) The optical backaction in the large amplitude regime where the
oscillator sweeps across the entire cavity resonance. The black line in (b,c)
indicates the position and linewidth of the cavity.
(d) Simulated ringup of a mechanical resonator with $\tilde \gamma_0
= 0.001$, coupled to a cavity with $\tilde \kappa = 100$,  $\tilde \Delta_0
= 30$, and $\com = 2000$.
From this timetrace the amplification rate $\tilde A^{-1}\tderl{\tilde A}{\tilde t}$ and
oscillation frequency $\tilde \omega_R$ are derived (e).\label{fig:overview}}
\end{figure}

Figure \ref{fig:overview}a shows a schematic of a Fabry-P\'erot
cavity where one of the mirrors can move, forming an
optomechanical resonator. The analysis presented here is not
limited to this particular system, but can be applied to any
cavity-optomechanical system including racetrack cavities
\cite{bagheri_natnano_high_amplitude}, photonic crystal
structures \cite{eichenfield_nature_optomechanical_crystals,
sun_NL_double_beam}, and even to microwave-cavity
optomechanical systems \cite{regal_natphys_cavity,
rocheleau_nature_lown, teufel_nature_strong_coupling}. All
these systems have in common that a displacement changes the
cavity frequency and thereby the number of photons inside it.
This results in a backaction on the mechanical element. In the
following, first the cavity-resonator dynamics will be
introduced in Sec. \ref{sec:dynamics}, followed by a study of
the appearance of self-sustained oscillations (Sec.
\ref{sec:sso}). Section \ref{sec:lar} shows what happens in the
large amplitude regime when the harmonic approximation breaks
down. Finally, Secs. \ref{sec:qe} and \ref{sec:duffing} study
the quantum efficiency and the role of mechanical
nonlinearities respectively.

\section{Cavity-resonator dynamics}
\label{sec:dynamics} The threshold for the onset of
optomechanically-induced SSO is well studied, both
experimentally and theoretically \cite{kippenberg_PRL_toroid,
lin_PRL_disks, kippenberg_OE_overview,
marquardt_physics_optomechanics_overview,
favero_natphot_overview, aguirregabiria_PRA_instability,
fabre_PRA_noise_reduction,
ludwig_NJP_optomechanical_instability}: For small harmonic
motion, the cavity field is only slightly perturbed and the
photon number oscillates at the mechanical frequency with an
in-phase and a quadrature part as illustrated in Fig.
\ref{fig:overview}b. In this linear regime the cavity response
is proportional to the motion amplitude. The quadrature part of
the oscillating photon occupation changes the damping rate from
its intrinsic value $\gamma_0$: For red detuned light the
optical backaction increases the total damping rate $\gamma$,
leading to cooling \cite{mancini_PRL_feedback,
marquardt_PRL_sideband_cooling,
wilson-rae_PRL_ground_state_cooling}. For blue detuning the
damping is reduced and can even become negative, the so-called
dynamic instability. In that case the amplitude grows
exponentially until it becomes limited by nonlinearities in
either the resonator or in the cavity, and the cavity dynamics
is strongly perturbed by the oscillator motion (Fig.
\ref{fig:overview}c). The main focus of this work is on the
role of the cavity, but the question of how mechanical
nonlinearities affect the maximum amplitude will be addressed
in Sec. \ref{sec:duffing}.

The coupled equations of motion for the displacement $u$ of a
harmonic oscillator and the optical field inside the cavity
$\sqrt{\hbar\Omega_L} a$ (in the frame rotating at the laser
frequency, $\Omega_L$), are \cite{fabre_PRA_noise_reduction,
kippenberg_OE_overview, marquardt_PRL_sideband_cooling}:
\begin{eqnarray}
m \ddot u & = & -m \omega_0^2 u - m \gamma_0 \dot u + \hbar \gom n \label{eq:u}\\
\dot a & = & -i(\Delta_0 + \gom u) a - \halfl \kappa a + \halfl \kappa \nmax^{1/2}\label{eq:a}.
\end{eqnarray}
Here, $\omega_0$ and $m$ are the resonator frequency and mass,
and $n = a^*a$ is the cavity photon number. The detuning is the
frequency difference between the laser and the cavity. For zero
displacement this is: $\Delta_0 = \Omega_L - \Omega_c|_{u =
0}$; a displacement changes this to $\Delta = \Delta_0 + \gom
u$, where $\gom \equiv -\pderl{\Omega_c}{u}$ is the
optomechanical coupling constant. When the detuning is zero,
the laser with power $P$ fills the cavity with photons until
the occupation reaches the steady-state value $\nmax =
4P\kappa_c/\kappa^2\hbar\Omega_L$, where the cavity linewidth
$\kappa$ is the sum of the external and intrinsic linewidths
($\kappa_c$ and $\kappa_i$ respectively). Note that for most
situations the linear coupling between the cavity frequency and
the displacement suffices, but refinements have been proposed.
These include a displacement-dependent $\gom$
\cite{bagheri_natnano_high_amplitude}, quadratic coupling
\cite{sankey_natphys_quadratic}, multiple optical resonances
\cite{metzger_PRL_instability,
pai_EPJD_instability_multi_resonances}, and effects due to
moving boundaries \cite{law_PRA_optomechanical_hamiltonian,
xuereb_njp_transfer_matrix}.

The full model (Eqs. \ref{eq:u} and \ref{eq:a}) contains 8
parameters, which are not all independent. By writing the model
in a dimensionless form
\begin{eqnarray}
\ddot {\tilde u} & = & - \tilde u - \tilde \gamma_0 \dot {\tilde u}  + c_{OM} \tilde a {\tilde a}^* \label{eq:uacc}\\
\dot {\tilde a} & = & -i(\tilde \Delta_0 + \tilde u) \tilde a - \halfl \tilde \kappa \tilde a + \halfl \tilde \kappa \label{eq:aacc},
\end{eqnarray}
the number of independent parameters is reduced to 4
\cite{marquardt_PRL_multistability}. Here, frequencies have
been normalized by $\omega_0$, and displacements by the
lengthscale $\omega_0/\gom$, so that $\tilde t = \omega_0 t$
and $\tilde u = \gom u /\omega_0$. Furthermore, $\tilde
\gamma_0^{-1} = \omega_0/\gamma_0$ is the intrinsic quality
factor, and the cavity field is scaled as $\tilde a =
a/\sqrt{\nmax}$ and $\tilde n = n /\nmax $. The coupling
strength is 
$\com = 2 \nmax \uzpm^2 \gom^2/\omega_0^2$, indicating that the
coupling can be viewed as the ratio of the
zero-point-motion-induced fluctuations of the cavity frequency
compared to the resonator frequency. Also note that the
optomechanical coupling coefficient $\gom$ appears in $\com$
squared due to the forward- and backaction
\cite{poot_physrep_quantum_regime}.

Equation \ref{eq:uacc} can be rewritten in terms of the complex
amplitude $\tilde U \equiv (\tilde u - i\dot {\tilde u})\exp(-i
\tilde t)$
\cite{heinrich_PRL_synchronization} by
discarding fast oscillating terms at frequencies $\sim
2\omega_0$ (the rotating wave approximation):
\begin{equation}
\dot {\tilde U} = - \half \tilde \gamma_0 \tilde U - i c_{OM} \tilde n_{\omega_0}, ~\textrm{where~} \tilde n_{\omega_0} = \langle \tilde n(t)e^{-i\tilde t} \rangle. \label{eq:Udot}
\end{equation}
This describes the evolution of the slowly varying amplitude
$\tilde A = |\tilde U|$ and phase $\theta = \angle \tilde U$.
$\tilde n_{\omega_0}$ is the Fourier component of the radiation
pressure at the oscillation frequency, indicating that only
frequencies near the resonance frequency contribute. In the
absence of coupling to the cavity (i.e., $\com = 0$) $\tilde A$
decays exponentially back to zero at a rate $\halfl \tilde
\gamma_0$. With coupling present, the out-of-phase part
$-c_{OM} \langle \tilde n \sin(\tilde t + \theta)\rangle$ of
the photon pressure changes the amplitude, whereas the in-phase
part $-c_{OM} \langle \tilde n \cos(\tilde t + \theta)\rangle$
changes the phase of the oscillations. As explained above, the
change in photon number is proportional to a small displacement
with some delay. This means that the cavity response can be
written as $\tilde n_{\omega_0} = \Phi \tilde U$, where the
complex response function $\Phi$ depends on $\tilde \Delta_0$
and $\tilde \kappa$. The real part of $\Phi$ is responsible for
the optical spring effect \cite{braginsky_PLA_instability,
sheard_PRA_opticalspring} as it changes $\theta$ at a constant
rate. On the other hand, the imaginary part modifies the
damping rate from $\tilde \gamma_0$ to $\tilde \gamma = \tilde
\gamma_0- 2c_{OM} \imag{\Phi}$.

\section{Self-sustained oscillations}
\label{sec:sso} From the discussion in the previous section, it
follows that when the imaginary part of $\Phi$ exceeds $\tilde
\gamma_0/2c_{OM}$ the total damping becomes negative and
oscillations start to grow exponentially as illustrated in Fig.
\ref{fig:overview}d. Ultimately, the oscillations are limited
in amplitude since the power provided by the cavity is finite
and the dynamics reaches a limit cycle
\cite{marquardt_PRL_multistability}. This can be understood as
follows: when the oscillations become too large, $\tilde
n_{\omega_0}$ is no longer linear in $\tilde U$ since the
cavity occupation is a nonlinear function of $u$, a Lorentzian
to be precise. Thus, when the oscillations exceed the range
where only the first order term of a Taylor expansion of
$\tilde n$ around $\tilde u = 0$ is needed (i.e. it exceeds the
dynamic range), the proportionality between $\tilde n_{\omega}$
and $\tilde U$ no longer holds, or, equivalently, $\Phi$
becomes amplitude dependent. An experimental signature of this
is the appearance of dips in the optical power coming out of
the cavity \cite{bagheri_natnano_high_amplitude,
anetsberger_natphys_nearfield, arcizet_nature_cavity,
carmon_PRL_instability, corbitt_PRA_opticalspring,
kippenberg_PRL_toroid, lin_PRL_disks} which results when the
oscillations sweep past the cavity resonance peak. In principle
$\Phi$ could depend on the complex amplitude at all past times
which would make the analysis of the system challenging.
However, since the optical cavity field adapts much faster
($\sim \kappa^{-1}$) than the resonator amplitude changes
($\sim \gamma_0^{-1}$), $\Phi$ only depends on the present
amplitude. Moreover, it is independent of the phase of the
oscillations, $\theta$. $\Phi$ is thus only a function of
$\tilde A$. Fig. \ref{fig:overview}e shows the evolution of the
amplification rate and the oscillation frequency, which are
directly related to the imaginary and real part of $\Phi$,
respectively. After the transient (Fig. \ref{fig:overview}d and
e) the oscillator reaches a steady state. The
backaction-limited amplitude of this limit cycle is a solution
to $\dot {\tilde A} = 0$ which is, according to Eq.
\ref{eq:Udot}, identical to $\imag{\Phi(\tilde A)} = \tilde
\gamma_0/2c_{OM}$. Note that because the absolute phase
$\theta$ is not fixed, thermal or backaction force noise will
lead to a slow phase diffusion of the oscillator
\cite{rodrigues_PRL_amplitude_noise}.

\begin{figure}[tb]
\includegraphics{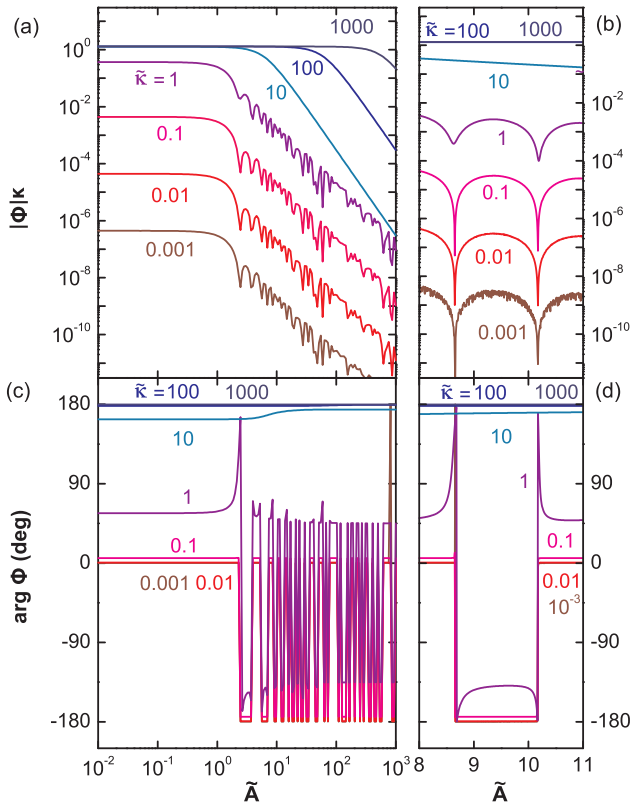}%
\caption{(Color online) Magnitude (a,b) and phase (c,d) of the steady state response of the cavity
photon number $n_\omega$ to a harmonic displacement with varying
amplitude $\tilde A$ for $\tilde \Delta_0 + \langle \tilde u
\rangle = 0.30\tilde \kappa$. The curves with different values for $\tilde \kappa$ are extracted from numerical time-domain simulations (App. \ref{app:numerics}).
The right panels show zooms of the area around
$\tilde A = 10$. \label{fig:amp}}
\end{figure}

The conclusion of the above discussion is that, to find the
maximum amplitude, one needs to know the function $\Phi(\tilde
A)$. When the backaction is not too strong and the amplitude is
not too high, the motion is to a good approximation harmonic.
However, as will be shown in Sec. \ref{sec:lar} for large
amplitudes this is no longer valid and the coupled
resonator-cavity dynamics will have to be calculated by
numerical integration. The cavity response for $\tilde u(\tilde
t) = \tilde A\cos(\tilde t)$ was calculated analytically by
Marquardt {\it et al.} \cite{marquardt_PRL_multistability} as
an infinite sum of Bessel functions:
\begin{equation}
\tilde a(\tilde t) = \frac{\tilde \kappa}{2} \sum_{m=-\infty}^{\infty} \frac{J_m(\tilde A)}{im + i\tilde\Delta_0 + \halfl \tilde\kappa} e^{im\tilde t - i \tilde A \sin(\tilde t)}, \label{eq:abessel}
\end{equation}
so that the amplitude-dependent cavity response becomes:
\begin{equation}
\tilde n_{\omega_0} = \left(\frac{\tilde \kappa}{2}\right)^2 \sum_{m=-\infty}^{\infty} \frac{J_m(\tilde A)J_{m-1}(\tilde A)}{(\tilde \Delta_0+m)^2 + (\halfl \tilde \kappa)^2- (\tilde \Delta_0+m) + \halfl i \tilde \kappa}. \label{eq:nbessel}
\end{equation}
Figure \ref{fig:amp} shows the amplitude-dependence of the
magnitude and phase of $\Phi = \tilde n_{\omega_0}/\tilde A$
for various values of $\tilde \kappa$ ranging from the deeply
resolved to the very unresolved sideband regimes. In both
regimes, a flat region exists on the left side of the plot
where the oscillations of $n$ are much smaller than $\nmax$ and
the cavity response is linear. By taking the limit $A
\rightarrow 0$ of Eq. \ref{eq:nbessel} the cavity response at
this plateau is obtained:
\begin{equation}
\Phi(\tilde A\rightarrow 0) = -\left(\frac{\tilde \kappa}{2}\right)^2 \frac{\tilde \Delta_0}{[\tilde \Delta_0^2 + (\halfl \tilde \kappa)^2 + \halfl i \tilde \kappa]^2-\tilde \Delta_0^2}.\label{eq:Philimit}
\end{equation}
In the unresolved sideband regime (USR) for large $\tilde
\kappa$ (and a constant ratio between $\tilde \kappa$ and
$\tilde \Delta_0$), $\Phi(\tilde A\rightarrow 0) \propto \tilde
\kappa^{-1}$. This is the reason that the plateaus in Fig.
\ref{fig:amp} collapsed onto a single curve for the USR. Fig.
\ref{fig:amp} also shows that the constant region extends up
to $\tilde A \sim \tilde \kappa$ in the USR; 
for larger amplitudes the cavity response rolls off smoothly as
$\tilde A^{-3}$.

In the resolved sideband regime (RSR) the situation is
different: still there is a constant region at low amplitudes,
but the magnitude of $\tilde \kappa \Phi$ now scales as $\tilde
\kappa^2$ which also follows from Eq. \ref{eq:Philimit}. This
can be understood as follows: the cavity linewidth acts as a
second-order low pass filter for $n(t)$ that filters out fast
oscillations. The slope of 20 dB/dec means that $\tilde A \sim
1$ is needed to have a response in $\tilde n \sim 1$. This
value of $\tilde A$ thus demarcates the end of the plateau as
shown in the Figure.
For larger amplitudes, $\Phi$ is, unlike in the USR, not a
smooth function of $\tilde A$. It contains many dips that
become deeper for smaller $\tilde \kappa$ as shown in Fig.
\ref{fig:amp}b. Here, only a few terms contribute to the sum in
Eq. \ref{eq:nbessel} which makes the oscillations in the
asymptotic form of $J_m(\tilde A) \rightarrow (2/\pi \tilde
A)^{1/2} \cos(\tilde A + m\pi/2 + \pi/4)$ for $\tilde A
\rightarrow \infty$ apparent. In contrast, in the USR many
terms contribute and the oscillations are washed out, resulting
in smooth curves.

Since the amplitude of the limit cycle is determined by the
condition that the total damping rate is zero ($\imag{\Phi} =
\tilde \gamma_0/2c_{OM}$, see above) the oscillations of
$\Phi(\tilde A)$ in the USR lead to a multitude of solutions
for the limit cycle of the oscillator for resolved
optomechanical systems \cite{marquardt_PRL_multistability},
whereas there is a unique solution for an unresolved system.
When expanding the equation of motion for $\tilde A$ (which is
obtained from Eq. \ref{eq:Udot} for small excursions $\delta
\tilde A$ around the limit cycle with amplitude $\bar A$) one
finds:
\begin{equation}
\dot{\delta \tilde A} = -\frac{\gamma_{\bar{A}}}{2}  \delta \tilde A, ~~ \gamma_{\bar{A}} = -2c_{OM} \left. \imag \pder{{\Phi}}{ \tilde A} \right |_{\tilde A = \bar{A}}
\end{equation}
For positive $\imag [\pderl{\Phi}{\bar{A}}]$ the limit cycle is
unstable and the excursions grow until reaching another
solution where both the damping rate is zero and where, at the
same time, the derivative is negative. Note that even though
the limit cycle itself is characterized by a vanishing damping
rate $\tilde \gamma(\bar{A}) = 0$, perturbations of the
oscillator amplitude are overdamped and return to $\bar{A}$ at
a rate $\gamma_{\tilde A}/2$. At the fixed point (the limit
cycle in the $u$-$\dot u$ plane corresponds to a fixed point in
the $\tilde U$ representation \cite{bagheri_synchronization})
the oscillation period is $\omega_0(1- \com \real \Phi)$, which
differs from that of the uncoupled harmonic oscillator due the
presence of the nonlinear optical potential
\cite{marquardt_PRL_multistability}.

\section{The large amplitude regime}
\label{sec:lar} In the discussion above it was assumed that the
motion of the oscillator is harmonic during the entire period.
This is a good approximation in the RSR, where the cavity
cannot respond quickly enough to the fast mechanical
oscillations and where the amplitudes are relatively small.
\begin{figure}[tb]
\includegraphics{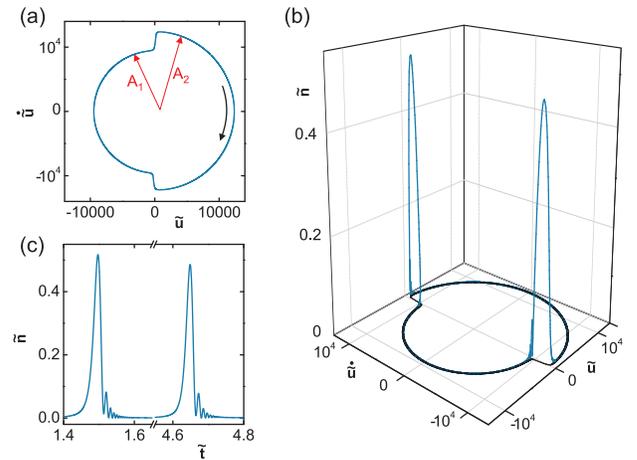}%
\caption{(Color online) Limit cycles of large amplitude motion in the USR for
$\tilde \gamma_0 = 0.01, \tilde \kappa = 100, \tilde \Delta_0 = 30$ and $c_{OM} = 2\cdot10^5$. (a)
Displacement-velocity phase portrait with the two amplitudes ($A_1$ and $A_2$) and the
direction of motion indicated. (b) Three dimensional phase portrait
with the photon number on the vertical axis. The black line shows the
projection onto the $\tilde u$-$\dot {\tilde u}$ plane. (c) Time trace of the
cavity occupation.\label{fig:portrait}}
\end{figure}

On the other hand, in the USR the amplitude will be shown to be
much larger and the cavity detuning oscillates back and forth
between large negative and positive values. There the photon
number thus changes rapidly from zero to a large number and
then back to zero. This kicks the resonator every time the
detuning crosses zero. When the resonator is moving forward
(i.e. to the right in Fig. \ref{fig:overview}a) work is done on
the resonator and the gained energy is transferred back to the
cavity during the backward motion. This is also reflected in
the dynamics of the oscillator. Instead of the ellipsoidal
shape for harmonic motion, Fig. \ref{fig:portrait}a shows a
mushroom-like phase portrait with two different amplitudes
($\tilde A_1$ and $\tilde A_2$). The step in the velocity
coincides with large peaks in $\tilde n$. Note that both these
effects were observed in a optomechanical system consisting of
a Bose-Einstein condensate inside a Fabry-P\'erot cavity
\cite{brennecke_science_optomechanics_BEC}. If the oscillation
would sweep through the resonance slowly, the cavity field
would always be in equilibrium with the input field and the
peaks would have the same height $\nmax$ since $n(t) =
\nmax/[\{2\Delta(u(t))/\kappa\}^2+1]$. The work done is then
$\hbar \gom \int^{\infty}_{-\infty} n(t) v(t) \intd t = \mp
\halfl \pi \nmax \hbar \kappa$, where the minus and plus signs
are for left and right moving resonators, respectively. Note
that the limits of the integral have been extended to $\pm
\infty$ since long before and long after the resonance is hit,
the cavity detuning is so large that the occupation is almost
zero at those times. Naturally, the work is proportional to the
laser power via $\nmax$ and to the linewidth: a larger $\kappa$
makes the Lorentzian lineshape wider and hence its area (to
which the work is proportional) larger. The work is, however,
independent of the optomechanical coupling coefficient since
the $\gom$-term in the force $\hbar \gom n(t)$ cancels the one
originating from the detuning $\Delta = \Delta_0 + \gom u$.
More importantly, since the contributions for the left and
right moving trajectories are equal, the net work during a
whole cycle is zero. Similar to the sideband formalism for
\emph{small} motion \cite{marquardt_PRL_sideband_cooling,
wilson-rae_PRL_ground_state_cooling, kippenberg_OE_overview} a
delay between the displacement and the cavity response is thus
also needed in the large amplitude regime to have the net
energy transfer required to overcome the mechanical damping.
One effect of such a delayed cavity response is shown in
\ref{fig:portrait}b and c. Interestingly, the timetrace of
$\tilde n(\tilde t)$ shows two different peak heights. When the
amplitude is small ($A_1$) the velocity is also low and the
cavity can fill up for a longer time than during the backward
motion with a larger amplitude ($A_2$) and a corresponding
larger velocity. Finally, the fast oscillations at the tail of
the peaks in Fig. \ref{fig:portrait}c result from interference
between the input field and the Doppler-shifted cavity photons
that entered the cavity at earlier times
\cite{marquardt_PRL_multistability, carmon_PRL_instability}.

\begin{figure}[tb]
\includegraphics{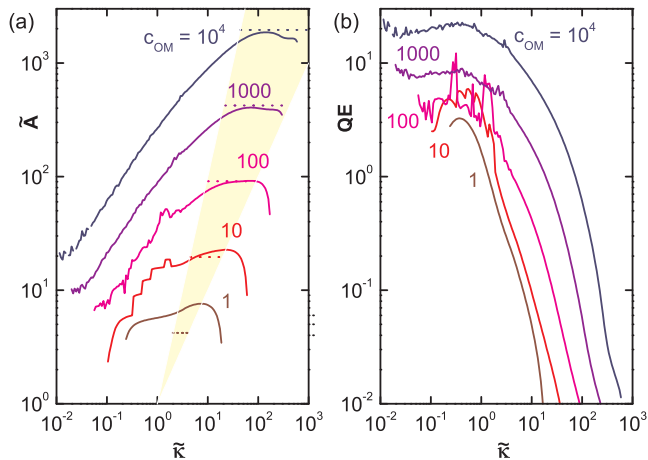}%
\caption{(Color online) Amplitude (a) and quantum efficiency (b) for $\tilde \gamma_0 =
0.01$ and $\tilde \Delta_0 = 3$, obtained by direct
numerical integration of Eqs. \ref{eq:uacc} and \ref{eq:aacc} (see App. \ref{app:numerics})
The dotted lines in (a) indicate the amplitude predicted by Eq.
\ref{eq:lar} and the shaded region indicates where
$\tilde A > \tilde \kappa$ and $\tilde A <
\tilde \kappa^2$ (see text). \label{fig:qe}}
\end{figure}

The amplitude of the limit cycle is thus determined by the
balance between the net energy gained during one period and the
intrinsic damping. Figure \ref{fig:qe}a shows the maximum
amplitude for different coupling strengths plotted against the
cavity linewidth. For low coupling the range over which the SSO
are stable is narrow, but this increases with increasing
$\com$. In the RSR the amplitude is rather small but increases
with increasing $\tilde \kappa$. The timetraces from which Fig.
\ref{fig:qe} is derived (for details see App.
\ref{app:numerics}) show that in the RSR for large coupling
$(\com \gtrsim 100)$ the system is chaotic
\cite{carmon_PRL_chaos}. The scatter of the curves in this
regime are reminiscent of this. Also note the multistability of
the amplitude in the curve for $\com = 10$
\cite{marquardt_PRL_multistability}. The amplitude keeps on
increasing with $\tilde \kappa$ while approaching the USR with
an exponent of about 2/3. Then far in the USR the maximum
occurs and is finally followed by sharp drop on the right. This
divides the region with stable SSO from the one where the
out-of-phase part of the backaction is too small to create a
negative damping rate.

From the simulations it is clear that the maximum amplitude
occurs in the flat region in the unresolved sideband regime.
There, the dynamics corresponds to that of Fig.
\ref{fig:portrait} and the first step to analyze the maximum
attainable amplitude is to consider the asymmetry in the peaks.
For a given initial amplitude, say $A_1$, the work done during
the kick can be calculated and from that the final amplitude
$A_2$ can be obtained by balancing the difference in work to
the mechanical damping
\cite{pai_EPJD_instability_multi_resonances}. However, even
though the peaks have different heights due to the velocity
difference, the net work done in one cycle is still zero when
accelerations \emph{during} the kick are not taken into
account. To proceed, first it is assumed that the kick happens
fast so that the displacements before and after are close,
which is the case when the time spent within the cavity
linewidth is small, i.e. $\tilde A_{1,2} \gg \tilde \kappa$.
Secondly, it is assumed that the detuning does not change too
fast so that $\ddot \Delta/\kappa^3$, $\dot \Delta^2/\kappa^4$
and higher order derivatives are small. This corresponds to
$\tilde A_{1,2} \ll \tilde \kappa^2$. The shaded area in Fig.
\ref{fig:qe}a shows that both conditions are satisfied for the
flat region near the maximum amplitude for the larger
couplings. The cavity response then becomes:
\begin{equation}
n(\tilde \Delta, \dot {\tilde \Delta}) \approx \frac{1}{1 + (2\tilde \Delta/\tilde\kappa)^2}\left[1 + \frac{32 \dot {\tilde \Delta}  \tilde \Delta/\tilde\kappa^3 }{\big [1 + (2\tilde \Delta/\tilde\kappa)^2 \big]^2}\right], \label{eq:larn}
\end{equation}
which has the usual Lorentzian shape for adiabatically slow
motion that does not yield a net energy transfer, but the
second term provides a correction for finite velocity. When
multiplying Eq. \ref{eq:larn} by $\dot {\tilde \Delta}$ and
integrating over time for both the forward and backward motion,
the net energy gain is found. Then by setting this equal to the
dissipated energy in one cycle the amplitude $A^2 =
\half(A_1^2+A_2^2)$ becomes:
\begin{equation}
A = \left(\frac{3}{4}\frac{\hbar^2\gom\nmax^2 Q_0}{m^2\omega_0^3}\right)^{1/3}~\mathrm{or}~\tilde A = \left(\frac{3}{4}\frac{c_{OM}^2}{\tilde \gamma_0}\right)^{1/3}. \label{eq:lar}
\end{equation}
The dotted lines in Fig. \ref{fig:qe}a show that the maximum
amplitude found from the numerical simulations is indeed given
by Eq. \ref{eq:lar} when both assumptions are satisfied. For
smaller coupling strength the shaded area is narrow and a
deviation between the simulated and calculated amplitude
develops, indicating that the assumptions break down. However,
the simulations show that the amplitude is larger than
predicted using Eq. \ref{eq:lar}, which thus still provides a
lower bound. Interestingly, Eq. \ref{eq:lar} shows that the
amplitude of the oscillations is independent of the cavity
decay rate. Although the work done during each kick is $\propto
\tilde \kappa$ the net energy transfered does not depend on
$\kappa$ since the latter requires a delay in the cavity
response, which is $\propto 1/\tilde \kappa$. These two
contributions thus balance each other. Also $\tilde A$ depends
only weakly on the mechanical damping rate $\tilde \gamma_0$
and on the coupling strength $\com$. Yet the largest amplitudes
are obtained for the strongest coupling and highest mechanical
quality factors. Also, although one needs blue detuning to
start the SSOs, the final amplitude is independent of the
detuning. This can be understood as follows: when the resonator
amplitude is so large that it rapidly sweeps over the cavity
and the decay rate is fast enough to empty the cavity before
the next kick (c.f. Fig. \ref{fig:portrait}) the exact time at
which the kicks happen (as set by the condition $u(t) =
-\Delta_0/\gom$) is not really important. Finally, note that
the 30 dB/dec slope of $\Phi$ shown in Fig. \ref{fig:amp}a is
reflected in the exponent of $1/3$ in Eq. \ref{eq:lar}.

\section{Quantum efficiency}
\label{sec:qe} In the previous section it was shown how to
reach the largest amplitudes, however this does not necessarily
mean that the process is efficient. The quantum efficiency $QE$
quantifies how many phonons are generated by a single photon.
Semiclassically, it is the ratio of the number of phonons
dissipated by the intrinsic damping $\gamma_0 \times \halfl m
\omega_0^2 A^2 / \hbar \omega_0$ and the rate of photons
entering the cavity $r_{in}$. Note that the total rate of
photons generated in the laser $P/\hbar\Omega_L$ is not used
since not all of these photons enter the cavity: this depends
on the ratio of $\kappa_i$ and $\kappa_c$. Since the quantum
efficiency is an intrinsic property of the cavity-resonator
system, only the photons that actually enter the cavity are
taken into account. Using Eq. \ref{eq:aacc} the rate equation
for the photon number is found: $\dot n = -\kappa n + r_{in}$,
with $r_{in}(t) = \kappa \nmax^{1/2} (a + a^*)/2$. The
time-dependence of $r_{in}$ enters via the non-trivial dynamics
of $a$. However, by averaging over one period of the mechanical
resonator and noting that for periodic motion $\langle \dot n
\rangle = 0$, the relation $\langle r_{in} \rangle = \kappa
\bar n$ between the average number of phonons in the cavity
$\bar n$ and the average rate is obtained. The quantum
efficiency then becomes:
\begin{equation}
QE = \frac{\gamma_0}{\kappa \overline
n}\left(\frac{A}{2\uzpm}\right)^2 = \frac{\tilde \gamma_0}{2 \tilde \kappa c_{OM}} \frac{\tilde A^2}{\bar{n}},
\end{equation}
where the term in the brackets is identified as the number of
phonons in the mechanical resonator. This shows that the larger
the amplitude becomes with a smaller average number of photons,
the larger $QE$.

From the simulated timetraces both the motion amplitude and the
average photon number are obtained. Figure \ref{fig:qe}b shows
the dependence of the simulated quantum efficiency for the data
shown in panel (a) on the linewidth and coupling strength. In
the deeply USR $QE$ is low because there the cavity can fill up
fast and $n$ is relatively large. The large amplitude reached
in this regime thus requires a large amount of photons and the
process is not efficient. When decreasing $\tilde \kappa$,
$\tilde A$ drops (Fig. \ref{fig:qe}a), but yet the quantum
efficiency goes up as $\bar{n}$ drops faster than $\tilde A^2$.
Finally, $QE$ saturates to a $\com$-dependent value above 1
when approaching the RSR.

\begin{figure}[tb]
\includegraphics{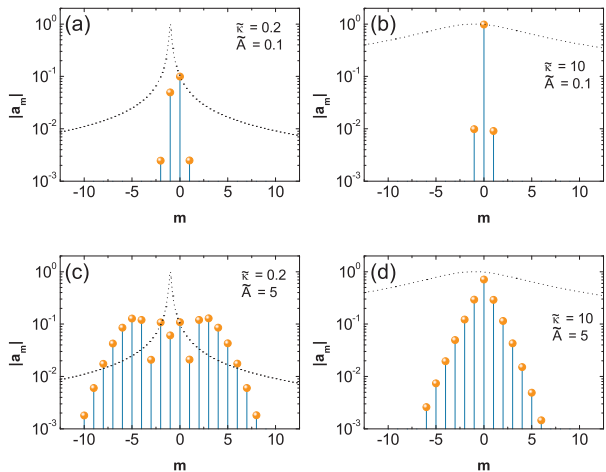}%
\caption{(Color online) Motion-induced sidebands of the optical field for small
(top) and large amplitude (bottom) in the RSR (left) and USR (right).
Negative $m$ corresponds to the emission of phonons, positive $m$ to
absorbtion. For $\tilde \Delta_0 = +1$ the cavity resonance is located at
$m = -1$ (dotted lines).
\label{fig:sidebands}}
\end{figure}

At first it might seem surprising that the quantum efficiency
in the RSR can exceed one. Note that this is not an artifact of
the definition of $r_{in}$: the same holds when using the total
laser power. The quantum efficiency is most easily understood
using the sideband picture of the optical backaction
\cite{kippenberg_science_optomechanics_overview}. Assume that
the laser is blue detuned from the cavity by $\sim \omega_0$.
Photons cannot enter the cavity since their energy is not
within the linewidth of the cavity. However, by emitting a
phonon the photons end up in the Stokes ($m = -1$) sideband and
have the right energy to enter the cavity. This seems to imply
that the QE can be at most 1 in the RSR. To understand why QE
can be much larger than unity, the sidebands (i.e. Fourier
coefficients) $a_m$ of $\tilde a(\tilde t)$ are analyzed by
expanding the $\exp(-i\tilde A \sin \tilde t)$ term in Eq.
\ref{eq:abessel}, resulting in:
\begin{equation}
a_m \equiv \int_{-\pi}^{\pi} \tilde a(\tilde t) e^{-im\tilde t} \intd \tilde t= \frac{\tilde \kappa}{2} \sum_{n=-\infty}^{\infty} \frac{J_n(\tilde A)J_{n-m}(\tilde A)}{in  + i\tilde \Delta_0 + \halfl \tilde \kappa}. \label{eq:ambessel}
\end{equation}
This shows that when $\tilde \Delta_0 \approx 1$ in the RSR a
pole occurs for $n = -1$. Now for small $\tilde A$ only $J_{0}$
and $J_{\pm 1}$ have an appreciable amplitude, and hence only
the $m=-1$ sideband becomes occupied as illustrated in Fig.
\ref{fig:sidebands}a. This sideband corresponds to the emission
of one phonon whereby the blue-detuned photon ends up at the
cavity resonance. In the USR more values of $n$ lie within the
cavity linewidth and both the $m = -1$ and $m=+1$ sidebands are
involved in the dynamics (Fig. \ref{fig:sidebands}b). The small
asymmetry between the Stokes and anti-Stokes sidebands provides
the energy to the resonator needed to overcome the intrinsic
damping. Also note the difference in amplitudes: In the RSR the
carrier ($m=0$) has a small amplitude as it is detuned from the
cavity and the $m = -1$ sideband is of the same order as the
carrier. In the USR the situation is reversed: Now the laser
light is within the cavity linewidth and hence the carrier is
of order unity, but the sidebands are much smaller. This
indicates that the phonon emission efficiency is lower in the
USR compared to the RSR.

For small amplitudes there are thus only one or two sidebands,
but for large amplitudes the situation is different (Fig.
\ref{fig:sidebands}c,d): now many more Bessel functions have a
nonvanishing value and more sidebands emerge. The number of
sidebands can be estimated as follows: the motion of the
oscillator frequency modulates the light inside the cavity up
and down. Light that entered the cavity at time $t_0$ becomes
frequency shifted by $\gom [u(t)-u(t_0)]$ at time $t$. The
number of available sidebands in the RSR is thus roughly equal
to the dimensionless amplitude $\tilde A$. Figure
\ref{fig:sidebands}c and d show that is indeed the case. In the
USR $a_m$ decays faster than in the RSR due to the shorter
lifetime. Finally, the overall quantum efficiency is the
balance between the number of available sidebands and the
probability to emit a single phonon. The sidebands for small
motion in Fig. \ref{fig:sidebands}a and b showed that the
latter process is very efficient in the RSR but less efficient
in the USR, although $QE$ can still exceed one in the USR for
large coupling. Combining all of this indicates that in the RSR
the quantum efficiency could be as high as $QE \sim \tilde A
\gg 1$, which is the result of multi-phonon emission. This
maximum value is indeed reached just before the end of the SSOs
on the left side of Fig. \ref{fig:qe}a and b.

\section{Optical vs mechanical nonlinearities}
\label{sec:duffing} 
\begin{figure}[tb]
\includegraphics{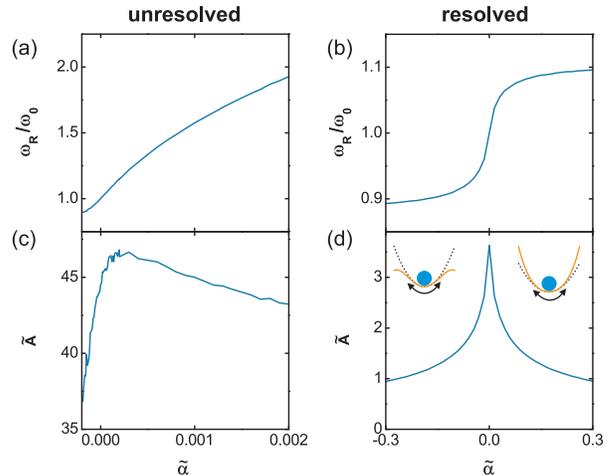}%
\caption{(Color online) Oscillation frequency (a,b) and amplitude (c,d) of nonlinear
resonators in the USR ($\tilde \kappa = 100$, left) and RSR ($\tilde \kappa = 0.01$, right) plotted against the nonlinearity parameter $\tilde \alpha$.
Both simulations have been done for $\tilde \gamma_0 = 10^{-5}$, $\tilde \Delta_0+\langle \tilde u
\rangle = 1$, and $c_{OM} = 1$. The insets show the shape of the mechanical potential energy. For negative $\tilde \alpha$ the springconstant weakens (left), and for  $\tilde \alpha > 0$ it stiffens (right).
\label{fig:nl}}
\end{figure}

So far only the optical nonlinearities that
limit the motion of the self-sustained oscillations have been
addressed. Another source of nonlinearity that can limit the
amplitude is mechanical in origin: when the motion is large,
stress induced in the resonator modifies the resonance
frequency from the small displacement value. For moderate
amplitudes, the equation of motion becomes that of the
well-known Duffing resonator, now combined with optomechanical
backaction:
\begin{equation} \ddot u = -\gamma_0 \dot u
-\omega_0^2 u - \omega_0^2 \alpha u^3 + \hbar\gom n/m.
\end{equation}
The nonlinearity parameter $\alpha$ has the units of
$\unit{m^{-2}}$ and the scaled version is defined as $\tilde
\alpha = \alpha\omega_0^2/\gom^2$. A positive value of $\tilde
\alpha$ indicates a stiffening mechanical spring, whereas a
negative value corresponds to a spring constant that decreases
with increasing amplitude. The most important effect of the
cubic term in this context is that it modifies the frequency of
the oscillations when the amplitude changes. For small $\tilde
\alpha$, the oscillator simply oscillates at the
optical-spring-modified resonance frequency. However, when
$\tilde \alpha$ is large the oscillation frequency shifts.
Figures \ref{fig:nl}a and b show the oscillation frequency of
two nonlinear resonators oscillating due to the optical
backaction of an unresolved (a) and a resolved optical cavity
(b). In the unresolved case, the oscillation frequency is
shifted by almost a factor of two for $\tilde \alpha = 0.002$
compared to the harmonic oscillator. For negative $\tilde
\alpha$ the frequency can be lowered to about 90\% of the
original value before the resonator escapes from the metastable
state around $u=0$ (see inset). Fig. \ref{fig:nl}c shows the
dependence of the motion amplitude on $\alpha$. Although there
is a dependence, only a $\sim 10\%$ change is visible. This is
because in the USR the backaction is broadband in nature, and a
change in resonance frequency will hardly affect the amplitude
of the oscillator. In the RSR the situation is different. Here
the laser should be close to the blue sideband of the cavity
(i.e., at $\Delta_0+ \gom \langle u \rangle = \omega_R$) to
have strong backaction heating. A change in the oscillation
frequency with amplitude will move the detuning away from this
optimal value and thus reduce the resulting oscillation
amplitude. The simulations in Fig. \ref{fig:nl}d show that this
is indeed the case. The oscillation amplitude is strongly
peaked around $\tilde \alpha = 0$ and drops when moving away
from this value. For large magnitudes of $\tilde \alpha$ the
feedback mechanism mentioned above clamps the amplitude to a
value of $\tilde A \sim 1$. Note that this can be used to
create a tunable optomechanical oscillator. First the nonlinear
resonator is set into oscillation with the optimal $\tilde
\Delta_0$, but then the detuning is slowly adjusted. Since the
amplitude is strongly peaked for $\omega_R = \Delta_0 + \gom
\langle u \rangle$ the oscillator  frequency can be dragged
along with the detuning, thus creating a widely tunable
oscillator.

Finally, it is also possible that the mechanical damping of the
resonator changes with amplitude as observed in the experiments
of Refs. \cite{zaitsev_IEEE_nl_damping,
zaitsev_nldyn_nl_damping,eichler_natnano_damping}. Since in
this case only the damping rate changes and $\omega_R$ remains
constant, this effect can easily be accounted for in the
present framework by self-consistently solving for the
amplitude using the \emph{average} mechanical damping rate.
However, as Eq. \ref{eq:lar} shows, the final amplitude depends
only weakly on the damping rate of the resonator, so the
influence of nonlinear damping on the maximum amplitude is
expected to be small.

\section{Conclusions}
The backaction limits on optomechanical motion have been
explored using the amplitude dependent cavity response at the
oscillation frequency. Above the threshold self-sustained
oscillations grow until the resonator reaches a limit cycle set
by the dynamic range of the cavity. The largest amplitudes are
obtained with an unresolved cavity, but the highest quantum
efficiencies are found in the resolved case. The latter can be
much larger than one because many sidebands are involved,
resulting in multi-phonon emission. For large amplitudes the
motion is anharmonic, and numerical simulations show that the
final amplitude is insensitive to the cavity detuning and
linewidth. Mechanical nonlinearities only have a modest effect
in the unresolved sideband regime, whereas in the resolved case
the shift in oscillation frequency due to anharmonicities can
have a strong effect on the amplitude.

\begin{acknowledgments}
M.P thanks the Netherlands Organization for Scientific Research
(NWO) / Marie Curie Cofund Action for support via a  Rubicon
fellowship. H.X.T. acknowledges support from a Packard
Fellowship in Science and Engineering and a career award from
National Science Foundation. This work was funded by the
DARPA/MTO ORCHID program through a grant from AFOSR. We thank
Florian Marquardt for discussions.
\end{acknowledgments}

\appendix
\section{Numerical simulations}
\label{app:numerics} Eqs. \ref{eq:u} and \ref{eq:a} are
integrated numerically using a fourth order Runge-Kutta
algorithm with adaptive step size, which gives the discretized
timetraces of the displacement, velocity, and (complex) cavity
amplitude as output. The transients where the oscillator
relaxes towards its final amplitude are discarded so that the
steady-state oscillations remain. The cavity response
$\Phi(\tilde A)$ (Fig. \ref{fig:amp}) is found by multiplying
the cavity occupation $n(t)$ by $\exp(-i\omega_0 t$ and
averaging over many cycles. For the data in Fig. \ref{fig:qe}
an initial state with a much larger amplitude was used to let
the oscillator relax preferentially to the solution with the
highest amplitude among the set of stable limit cycles. The
amplitude $\tilde A$ is obtained by calculating the
root-mean-squared value of the displacement and multiplying by
$\sqrt{2}$. This ensures that the obtained value equals the
amplitude for sinusoidal oscillations. An alternative
definition where half the difference between the minimum and
maximum value of $u(t)$ is used has almost identical numerical
values, but naturally shows more scatter near the chaotic
regions.



\end{document}